\documentclass[twocolumn,showpacs,preprintnumbers,amsmath,amssymb]{revtex4}
\usepackage{graphicx}% Include figure files
\usepackage{dcolumn}% Align table columns on decimal point
\usepackage{bm}% bold math

\begin{document}

%\preprint{APS/123-QED}
\newif\ifplot
\plottrue
%\plotfalse
\newcommand{\RR}[1]{[#1]}
\newcommand{\intsum}{\sum \kern -15pt \int}
\newfont{\Yfont}{cmti10 scaled 2074}
\newcommand{\Y}{\hbox{{\Yfont y}\phantom.}}
\def\O{{\cal O}}
\newcommand{\bra}[1]{\left< #1 \right| }
\newcommand{\braa}[1]{\left. \left< #1 \right| \right| }
\def\Bra#1#2{{\mbox{\vphantom{$\left< #2 \right|$}}}_{#1}
\kern -2.5pt \left< #2 \right| }
\def\Braa#1#2{{\mbox{\vphantom{$\left< #2 \right|$}}}_{#1}
\kern -2.5pt \left. \left< #2 \right| \right| }
\newcommand{\ket}[1]{\left| #1 \right> }
\newcommand{\kett}[1]{\left| \left| #1 \right> \right.}
\newcommand{\scal}[2]{\left< #1 \left| \mbox{\vphantom{$\left< #1 #2 \right|$}}
\right. #2 \right> }
\def\Scal#1#2#3{{\mbox{\vphantom{$\left<#2#3\right|$}}}_{#1}
%\kern -2pt
{\left< #2 \left| \mbox{\vphantom{$\left<#2#3\right|$}}
\right. #3 \right> }}

\title{
Three Nucleon Force Effects in Intermediate Energy Deuteron Analyzing
Powers for \mbox{\boldmath$\mathnormal{dp}$} Elastic Scattering
}
% Force line breaks with \\

\author{K.\ Sekiguchi$^{1}$}
\email{kimiko@lambda.phys.tohoku.ac.jp}
\author{H.\ Okamura$^{2}$}
\thanks{Deceased.}
\author{N.\ Sakamoto$^{3}$}
\author{H.\ Suzuki$^{4}$}
\author{M.\ Dozono$^{5}$}
\author{Y.\ Maeda$^{6}$}
\author{T.\ Saito$^{6}$}
\author{S.\ Sakaguchi$^{3}$}
\author{H.\ Sakai$^{3,7}$}
\author{M.\ Sasano$^{3}$}
\author{Y.\ Shimizu$^{8}$}
\author{T.\ Wakasa$^{5}$}
\author{K.\ Yako$^{7}$}
\author{H.\ Wita{\l}a$^{9}$}
\author{W.\ Gl\"ockle$^{10}$}
\author{J.\ Golak$^{9}$}
\author{H.\ Kamada$^{11}$} 
\author{A.\ Nogga$^{12}$}
\affiliation{
$^{1}$Department of Physics, Tohoku University, Sendai, 980-8578, Japan}
\affiliation{$^{2}$Research Center for Nuclear Physics, Osaka
University, Ibaraki, Osaka, 567-0047, Japan}
\affiliation{
$^{3}$RIKEN Nishina Center, Wako, 351-0198, Japan}
\affiliation{
$^{4}$Graduate School of Pure and Applied Sciences, 
University of Tsukuba, Tsukuba, 305-8571, Japan}
\affiliation{$^{5}$Department of Physics, Kyushu University,
Fukuoka, 812-8581, Japan}
\affiliation{
$^{6}$Faculty of Engineering, University of Miyazaki, 
Miyazaki, 889-2192, Japan}
\affiliation{
$^{7}$Department of Physics, University of Tokyo, Tokyo, 113-0033, Japan}
\affiliation{
$^{8}$Center for Nuclear Study, University of Tokyo, Tokyo, 113-0033, Japan}
\affiliation{
$^{9}$Institute of Physics, Jagiellonian University,
PL-30059 Cracow, Poland}
\affiliation{
$^{10}$Institut f\"ur theoretische Physik II,
Ruhr-Universit\"at Bochum, D-44780 Bochum, Germany}
\affiliation{
$^{11}$Department of Physics, Kyushu Institute of Technology, 
Kitakyushu 804-8550, Japan}
\affiliation{$^{12}$
Institut f\"ur Kernphysik, Institute for Advanced Simulation
and J\"ulich Center for Hadron Physics, D-52428 J\"ulich, Germany}

\date{\today}% It is always \today, today,
             %  but any date may be explicitly specified

\begin{abstract}
A complete high precision set of deuteron analyzing powers for
elastic deuteron-proton ($dp$)  
scattering at 250 MeV/nucleon (MeV/N) has been measured.
The new data are presented together with data from previous
measurements at 70, 100, 135 and 
200 MeV/N.
They are compared with the results of three-nucleon (3N) 
Faddeev calculations based on 
 modern nucleon-nucleon (NN) potentials alone or combined 
with two models of three nucleon forces (3NFs):  
the Tucson-Melbourne 99 (TM99) and Urbana IX. 
At 250 MeV/N large discrepancies between pure NN models and data, 
which are not resolved by including 3NFs, 
were found at c.m.\ backward angles of $\theta_{\rm c.m.}\gtrsim 120^\circ$
for almost all the deuteron analyzing powers. 
These discrepancies are  quite similar to those found for the
cross section at the same energy. 
We found small relativistic effects that cannot resolve the 
discrepancies with the data indicating that other, short-ranged 3NFs
are required to obtain a proper description of the data.
\end{abstract}

% PACS, the Physics and Astronomy
                             % Classification Scheme.
%\keywords{Suggested keywords}%Use showkeys class option if keyword
                              %display desired
\maketitle

A hot topic in present day few-nucleon system studies is 
exploring the properties of three-nucleon forces (3NFs) 
that appear when more than two nucleons ($A \ge 3$) interact.
The 3NFs arise naturally in the standard meson exchange picture
in which the main ingredient  is considered to be 
$2\pi$-exchange between three nucleons along with 
the $\Delta$--isobar excitation, the mechanism initially proposed 
a half century ago by Fujita and Miyazawa~\cite{fuj57}.
Further enhancements have led to 
the Tucson--Melbourne (TM)~\cite{TM} and  Urbana IX 3NF~\cite{uIX}
models. 
New impetus to study 3NFs has come from 
chiral effective field theory ($\chi$EFT)
descriptions of nuclear interactions. 
In that framework consistent two-, three-, and many-nucleon 
forces are derived on the same footing~\cite{kolk1994,epel2006}. 
The first non-zero contribution to 3NFs appears  in $\chi$EFT 
at the next-to-next-to-leading order (N$^2$LO) of the chiral expansion.  
That explains why 3NFs are relatively small compared to 
NN forces  (2NFs) and why their effects are easily masked. 
Therefore, it is, in general, hard to find evidence for them.

The first evidence for a 3NF was found
in the three-nucleon bound states, $^3\rm H$ and $^3 \rm He$
\cite{chen,sasakawa}.
The binding energies of these nuclei are not reproduced  by 
exact solutions of three-nucleon 
Faddeev equations employing modern NN 
forces only, {\it i.e.}\ AV18~\cite{AV18}, 
CD Bonn~\cite{cdb}, Nijmegen I, II 
and 93~\cite{nijm}.  
The underbinding of $^3$H and $^3$He can be 
 explained by adding a 3NF, mostly based on $2\pi$-exchange, acting  
between three nucleons~\cite{chen,sasakawa,underbind}.
The importance of 3NFs has  been further supported 
by the binding energies of light mass nuclei. 
{\it Ab initio} 
microscopic calculations of light mass nuclei, such as 
Green's Function Monte Carlo~\cite{piep2002}
and no-core shell model calculations~\cite{navr03}, 
highlight the necessity of including 3NFs
to explain the binding energies and low-lying levels of these nuclei.

In addition to the study of 3N bound states, 
three nucleon scattering  is an attractive candidate 
for a further, more detailed investigation of properties of 3NFs, such as 
their momentum, spin, and/or isospin dependence. 
In elastic nucleon-deuteron ($Nd$) scattering and in deuteron breakup
reactions one can measure 
not only differential cross sections 
but also a rich set of polarization observables under various 
kinematic conditions.  
The importance of 3NFs in $Nd$ elastic scattering
was shown for the first time in \cite{wit98}.
Clear signals from 3NFs were found around the cross section 
minimum
occurring at c.m. angle $\theta_{c.m.} \approx 120^{\circ}$  
for incident energies above $60~\rm MeV/N$.
Since then experimental 
studies of proton--deuteron ($pd$) and/or 
neutron--deuteron ($nd$) elastic scattering 
at intermediate energies have been performed
extensively and have provided precise data for cross sections
\cite{sak96,sak2000,sek2002,hatanaka02,erm2005,sek2005,maeda06} 
and spin observables, such as 
analyzing powers
\cite{sak96,sak2000,sek2002,hatanaka02,erm2005,maeda06,anpow1,anpow2,Przewoski_PRC_74_2006_064003}, 
spin correlation coefficients~\cite{Przewoski_PRC_74_2006_064003}, 
and polarization transfer coefficients~\cite{hatanaka02,sek2004,hamid2007}.
Large discrepancies between data and 
rigorous Faddeev calculations with modern NN forces alone have been 
reported, which are particularly significant in the 
angular region of the cross section  minimum and at 
incident nucleon energies above about 60 MeV/N.
In the case of  the cross sections,
a larger part of the discrepancies are removed 
combining  the NN forces with a 3NF such as the TM99~\cite{tm99} or Urbana IX.
Those  results can be taken as  clear 
signatures for 3NF effects in $Nd$ elastic scattering and the data
themselves 
form a solid basis to test new theories.
In contrast to the elastic scattering cross section the theoretical 
predictions with 3NFs included still 
have difficulties in reproducing data for some spin observables.

In Refs.~\cite{hatanaka02,maeda06},
precise data for the cross section and 
nucleon analyzing powers at 250 MeV/N 
in $pd$ and $nd$ elastic scattering were reported. 
Contrary to the results obtained at lower energies, 
%e.g. 135 MeV/N,
the discrepancies in the cross sections were only partially removed 
by incorporating 3NFs
and large discrepancies of $\approx 40 \%$ still remained 
 at backward angles 
$\theta_{\rm c.m.} \gtrsim 120^\circ$.

The results of Refs.~\cite{hatanaka02,maeda06}
present a new challenge to be solved and show that 
an energy dependent study of a variety of spin observables 
should be a good source to obtain a consistent understanding
of the dynamics of nuclear forces in elastic $Nd$ scattering.
Especially the deuteron tensor analyzing powers
should be interesting,  because matrix elements of
the 2$\pi$-exchange 3NF consist of momentum-spin coupled 
operators.
Data for the cross section and spin observables 
in elastic $pd$/$nd$ scattering have been accumulated 
at incident energies $\lesssim 200~\rm MeV/N$.
However, only a few precise data exist around 200--300 MeV/N.
The new accelerator facility at RIKEN, RI beam factory (RIBF),
where polarized deuteron beams are available up to 440
MeV/N, offers opportunities to make these experimental studies.
As the first experiment with a vector and tensor 
polarized deuteron beam at RIBF,
the measurement of elastic $dp$ scattering at 250 MeV/N
was performed, providing a complete set of data for 
all deuteron analyzing powers $iT_{11}$, $T_{20}$, $T_{21}$ and $T_{22}$, 
in a wide angular range $\theta_{\rm c.m.} = 36^\circ$--$162^\circ$.

At RIBF 
the vector and tensor polarized deuteron beam  
was accelerated first by the injector cyclotrons
AVF and RRC up to 90 MeV/N, and then 
up to 250 MeV/N by the new superconducting cyclotron SRC.
The measurement of elastic $dp$ scattering was performed with 
the new beam line polarimeter BigDpol, installed
at the extraction beam line of the SRC. 
A polyethylene ($\rm CH_2$) target with 
a thickness of $\rm 330~mg/cm^2$ was used as a hydrogen target.
In the BigDpol four pairs of plastic scintillators coupled with
photo-multiplier tubes were mounted in two independent planes,
$90^\circ$ apart in azimuthal angle
and operated in kinematic coincidence for elastic $dp$ scattering. 
The opening angle of the BigDpol was $8^\circ$--$70^\circ$.
The solid angles were determined by the proton detectors
and the angular spread was $\Delta\theta_{\rm lab.}\pm 1^\circ$.
For subtraction of events 
from the proton knock-out reaction on carbon nuclei
${\rm ^{12}C}(d,dp){\rm ^{11}B}$
a measurement with a graphite target was also performed.

The data were taken with polarized and unpolarized beams for
different combinations of the incoming polarization  
given in terms of the theoretical maximum polarization values
as $(p_{Z},p_{ZZ})=(0,0)$, $(1/3,-1)$, $(-2/3,0)$ and $(1/3,1)$.
Those polarization modes were changed cyclically at intervals
of 5 seconds by switching the RF transition units of the polarized 
ion source.
The beam polarizations were monitored continuously 
with a beam line polarimeter prior to acceleration by the SRC 
using elastic $dp$ scattering at 90 MeV/N. 
The analyzing powers for this reaction were calibrated 
in the previous measurement 
by using the $^{12}{\rm C}(d,\alpha)^{10}{\rm B}^{*}\left[2^{+}\right]$ 
reaction, the $A_{zz}\left(0^\circ\right)$ of which 
is exactly $-1$ because of parity conservation~\cite{Hossein_EPJA41_2007_383}.
In the present measurement
typical values of the beam polarizations were 80\% of the 
theoretical maximum values.

The deuteron analyzing powers for elastic $dp$ scattering
are expressed through the unpolarized ($\sigma_0$) and polarized 
($\sigma$) cross sections together with the vector and tensor polarizations of 
the incoming deuteron ($p_Z$ and $p_{ZZ}$) as 
\begin{eqnarray}
%\label{poldcs_aij}
\sigma &=& \sigma_0 \biggl\{ 1
      + \sqrt{3}\,iT_{11}\left(\theta\right) p_Z \sin \beta \cos\phi 
\nonumber \\
&& \hspace*{2.2em}{} + \frac{1}{\sqrt{8}}\,T_{20}\left(\theta\right)p_{ZZ} \left(3\cos ^2 \beta -1\right) 
\nonumber \\ 
&& \hspace*{2.2em}{} + \sqrt{3}~T_{21}\left(\theta\right) p_{ZZ} \cos\beta \sin \beta \sin \phi
\nonumber \\ 
&& \hspace*{2.2em}{} + \frac{\sqrt{3}}{2}\,T_{22}\left(\theta\right)p_{ZZ} \sin ^2\beta \cos 2\phi 
\biggr\}
\;,
\label{poldcs_aij}
\end{eqnarray}
where $\theta$ and $\phi$ are the polar and azimuthal scattering angles,
respectively \cite{ohl72}. 
The $\beta$ is defined as the angle between the spin direction and 
the beam direction. 
In this experiment the polarization axis of the deuteron beam
was rotated with a Wien filter system to the direction 
required for the measurement of a particular analyzing power~\cite{okamura95}.
For the measurement of the analyzing powers $iT_{11}$, $T_{20}$, and 
$T_{22}$
the polarization axis was normal to the horizontal plane.
For the $T_{21}$ measurement
the spin symmetry axis was rotated in the reaction plane 
and aligned at an angle $\beta = 38.0^\circ \pm 0.51^\circ$ 
to the beam direction.

\begin{figure}[htbp]
\centering
\includegraphics[scale=0.4,angle=0]{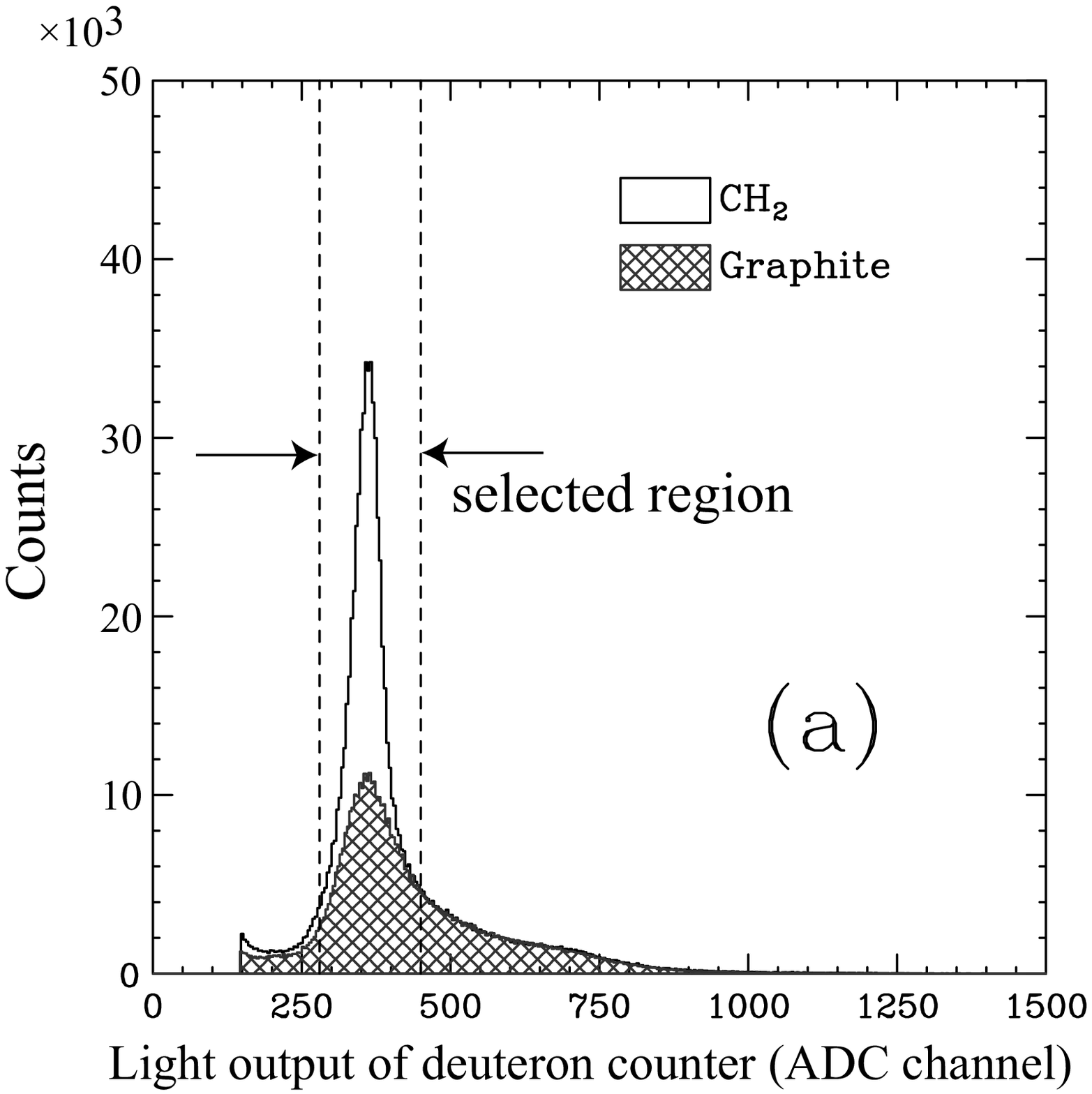}\\
\includegraphics[scale=0.4,angle=0]{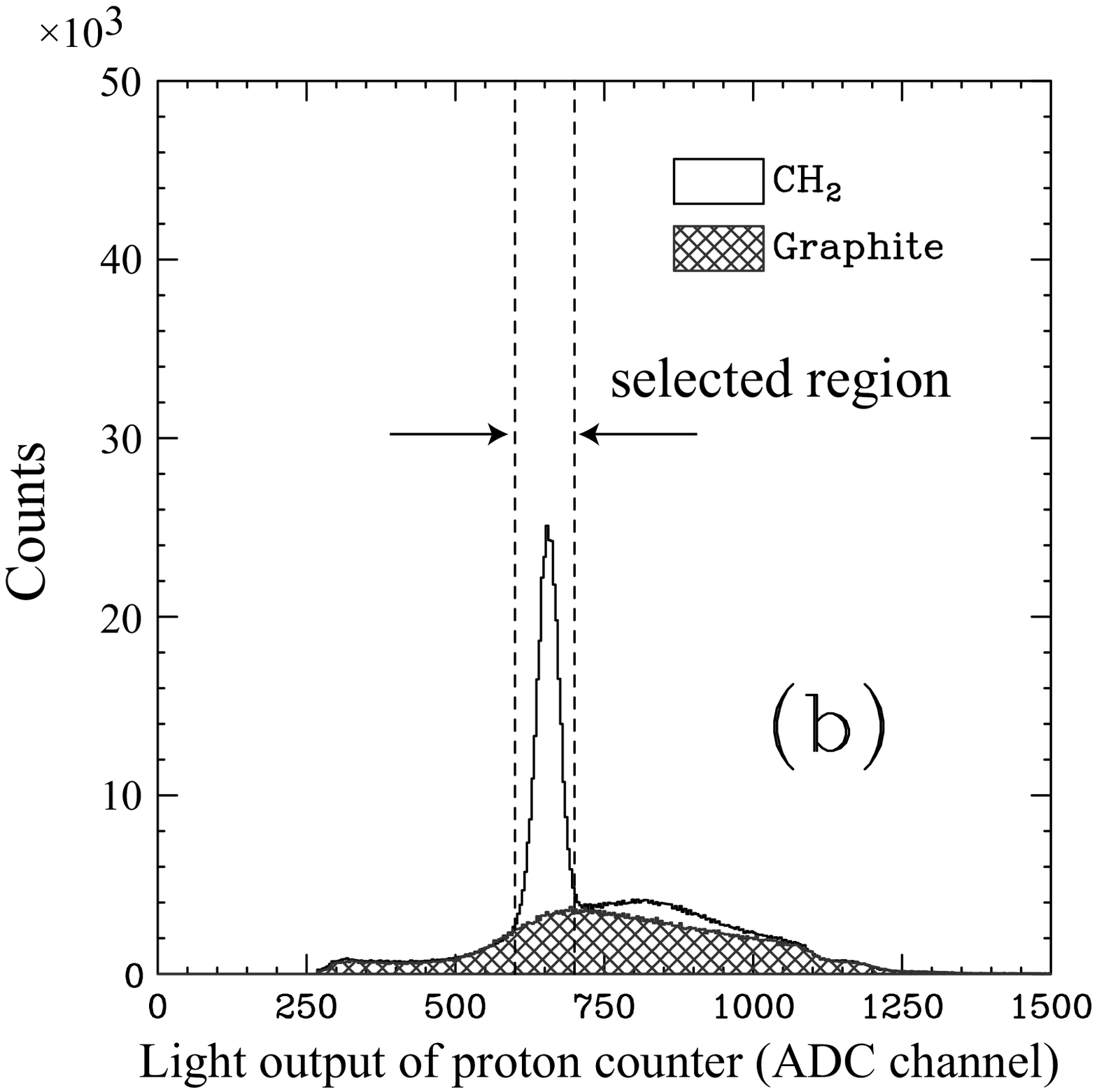}\\
\includegraphics[scale=0.4,angle=0]{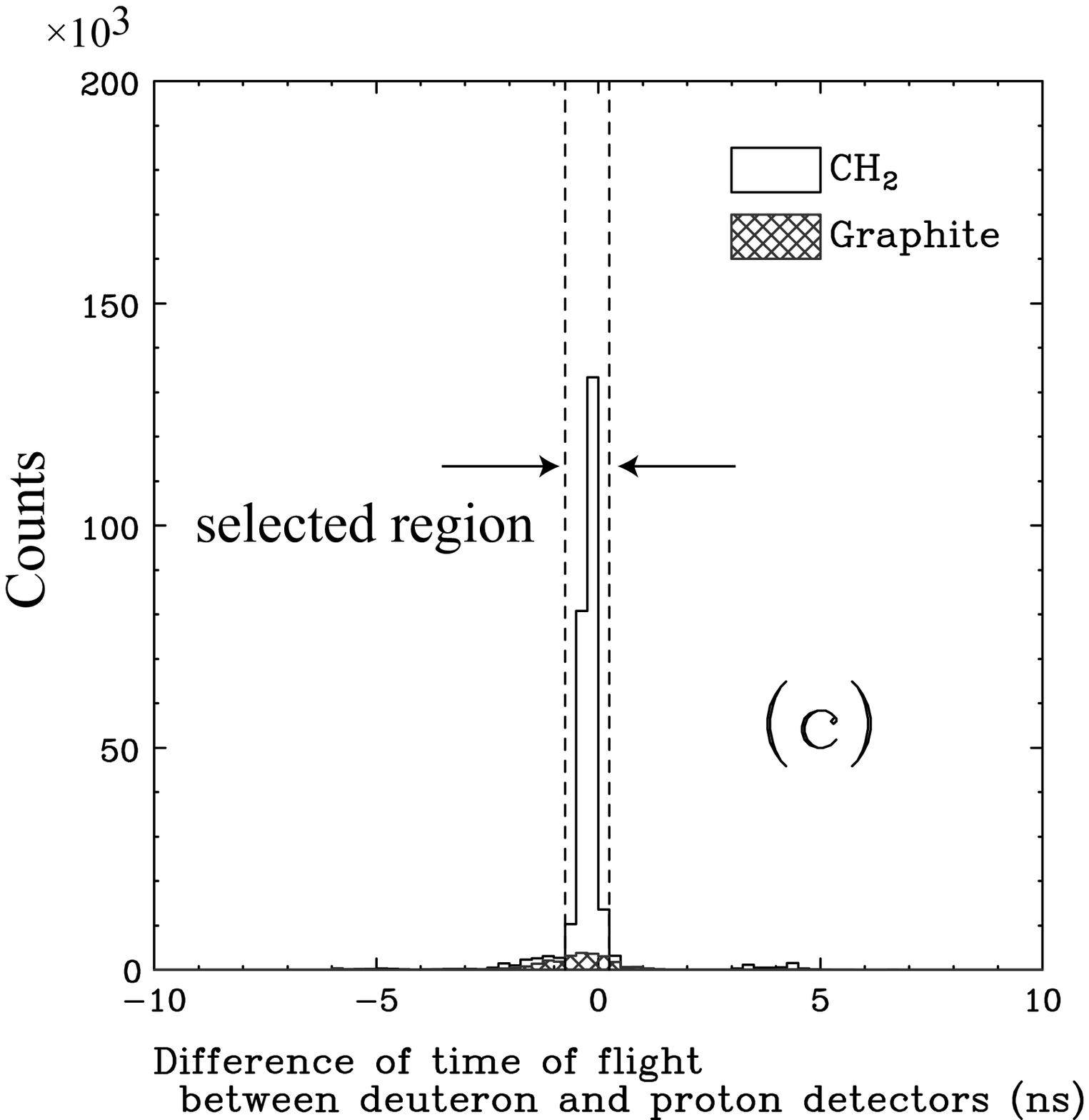}
\caption{Spectra of light outputs of of the scintillation
counters for the scattered deuterons (a) and the recoil 
protons (b). The deuteron and proton detectors 
were placed at $19.5^\circ$ and $55^\circ$, respectively.
After selecting events in (a) and (b) 
the spectra of time of flight difference between the 
deuteron and proton detectors was obtained (c).
}
\label{spectra}
\end{figure}

Figures~\ref{spectra}(a) and (b) 
show spectra of scintillator light outputs 
of one pair of the deuteron and proton detectors
placed at $19.5^\circ$ and $55^\circ$, respectively.
The spectra obtained with the $\rm CH_2$ target
together with the graphite target are shown.
Impinging deuterons and protons punched through 
the detectors.
Identification of the scattered deuterons and 
recoil protons 
was performed 
by cuts for the spectrum with the $\rm CH_2$ target
as shown in Figs.~\ref{spectra}(a) and (b).
After the selection of events
a spectrum of the time-of-flight difference 
between the deuteron and proton detectors 
was obtained (see Fig.~\ref{spectra}(c)).
As for asymmetry determination of elastic $dp$ scattering
the selected events within the gate shown in Fig.~\ref{spectra}(c) 
were used. 
Since the contribution from carbon nuclei was small 
in Fig.~\ref{spectra}(c)
the background subtraction was not performed in the analysis.

\begin{figure*}[htbp]
\centering
\includegraphics[scale=0.37,angle=0]{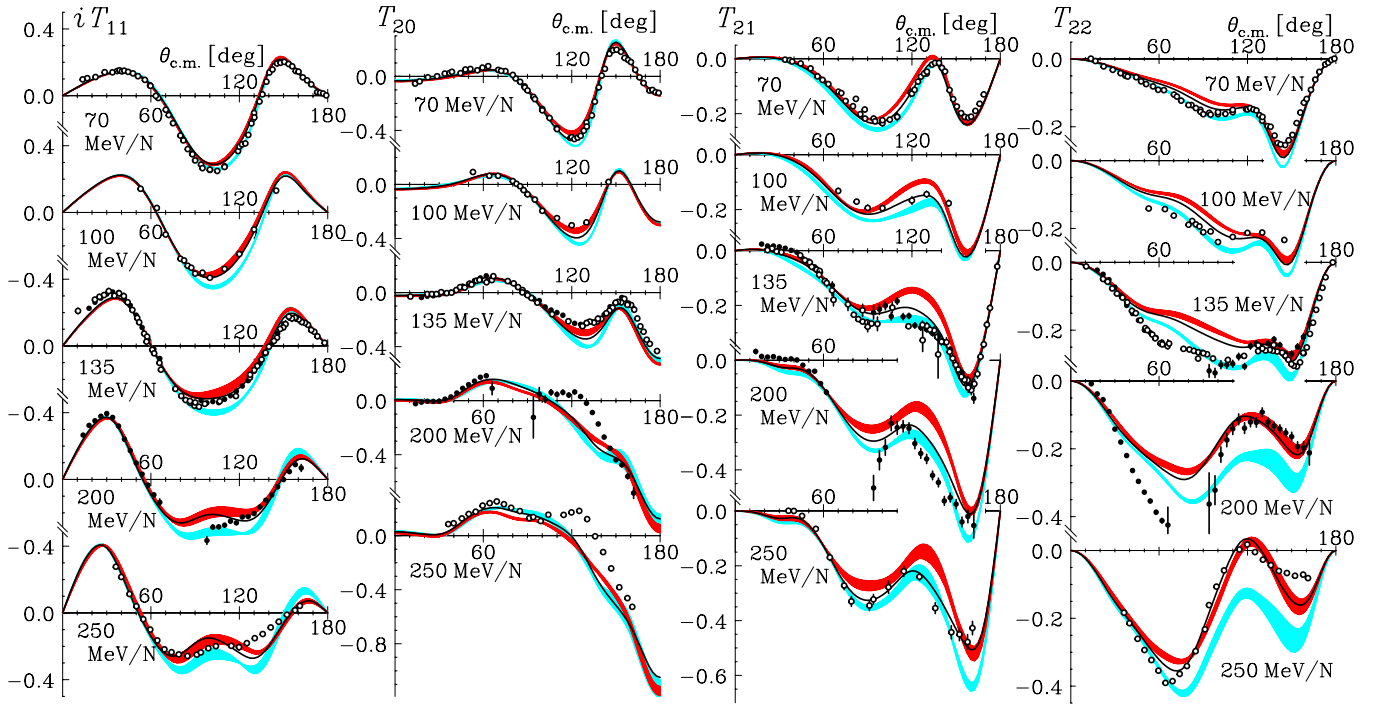}
\caption{(Color online) The deuteron analyzing powers $iT_{11}$,
  $T_{20}$, $T_{21}$, and  $T_{22}$ for $dp$ elastic 
scattering at 70, 100, 135, 200, and 250 MeV/N. The light shaded
(blue) bands contain predictions of modern NN potentials: AV18, CD Bonn,
Nijmegen I and II. The dark shaded (red) bands result when those potentials
are combined with TM99 3NF, properly adjusted to reproduce the $^3$H
binding energy. The solid line is the result obtained with the combination
AV18+Urbana IX. The pd data are: at 70 MeV/N (open circles) from
Ref.~\cite{sek2004}, at 100 MeV/N (open circles) from
Ref.~\cite{sek2002},  at 135 MeV/N (open circles) from
Ref.~\cite{sek2002} and (solid circles) 
from Ref.~\cite{Przewoski_PRC_74_2006_064003}, 
 at 200 MeV/N (solid circles) from
Ref.~\cite{Przewoski_PRC_74_2006_064003}, and at 250 MeV/N 
(open circles) from the present study. 
}
\label{tij_dp}
\end{figure*}

We  present these new data and compare them with 
theoretical predictions based on different dynamical input in 
Fig.~\ref{tij_dp}.  The statistical uncertainties, which are less than 0.01 
for all the deuteron analyzing powers, are also shown.
In order to clarify the energy dependence, 
the previously measured  deuteron analyzing powers 
at 70, 100 and 135 MeV/N~\cite{sak96,sak2000,sek2002,sek2004}
and at 200 MeV/N~\cite{Przewoski_PRC_74_2006_064003} are
also included together with theoretical predictions. 
The deuteron analyzing powers at 70, 100, 135 MeV/N 
have been reported 
in Refs.~\cite{sak96,sak2000,sek2002,sek2004} in the Cartesian
notation. Here  we show them in the spherical description.
 The theoretical predictions shown are based on modern NN forces alone
and on their
combinations with 3NF models~\cite{glo96,hub97}. We used high precision NN
potentials AV18, CD Bonn, and Nijmegen I and II alone [light shaded
(blue) bands in Fig.~\ref{tij_dp}] or combined them with the
TM99 3NF~\cite{tm99} with the cutoff $\Lambda$ values that yield,
for a particular NN force and TM99 combination, a reproduction
of the $^3$H binding energy [dark shaded (red) bands in
Fig.~\ref{tij_dp}]. 
In case of the AV18 NN force, 
we also combined it with the  Urbana IX 3NF~\cite{uIX} 
(solid lines in Fig.~\ref{tij_dp}).  
In our calculations we neglected the Coulomb force acting between two
protons. At energies considered here its effect on polarization
observables is small~\cite{deltuva2005}.

For the vector analyzing power $iT_{11}$ 
the predicted 3NF effects at 70 MeV/N are small and the data are
satisfactorily described, independently of whether 3NFs are included. 
At higher energies, however, a discrepancy between
data and theory based on NN forces only is clearly seen for
c.m. angles  $\theta_{\rm c.m.} \gtrsim 80^\circ$.
%, which increases with energy. 
At 100 and 135 MeV/N adding TM99 or Urbana IX 3NF provides 
quite a good description of the data. 
That is not the case at 200 and 250 MeV/N, where only a part of 
the discrepancy is
removed and the full theory is quite far from data in some angular
regions. 
For  $\theta_{\rm c.m.} \lesssim 60^\circ$ at all energies 3NF effects 
are negligible and data are reproduced by pure NN theory. 

For the tensor analyzing power $T_{20}$, 
the predicted 3NF effects are smaller than for $iT_{11}$. 
Again, at 70 MeV/N the theory
reasonably well reproduces the data and at 100 and 135 MeV/N
adding 3NFs brings the theory closer to the data. At 200 and 250
MeV/N, where the predicted 3NF effects for $T_{20}$ are rather
moderate, drastic discrepancies between theory and data in the angular
region  $80^\circ \lesssim \theta_{\rm c.m.} \lesssim 140^\circ$ 
are clearly seen, which are not explained by adding standard 3NFs. 

The data for the tensor analyzing power $T_{21}$ are reproduced quite
well by calculations with the 2NFs only for all incident energies. The
large effects of the TM99 3NF predicted for energies $ \gtrsim 100$~MeV are
not supported by the $T_{21}$ data. They differ clearly from the
smaller effects of the Urbana IX 3NF, which generally leads to a better
description for that observable. 

For the tensor analyzing power $T_{22}$, 
the discrepancies between the data and the predictions
based on 2NFs only become larger 
in magnitude and expand to the backward angles 
with increasing incident energy. 
For that observable the predicted 3NF effects are especially 
large and similar in magnitude for the 2NFs plus TM99 and 
Urbana IX models. 
At 200 MeV/N and less
the predictions taking into account 3NFs 
have good agreement to the data 
at the backward angles,
however the data in the angular region
$40^\circ \lesssim \theta_{\rm c.m.} \lesssim 120^\circ$
are not described by any theoretical predictions.
At 250 MeV/N the overall agreement 
is improved by taking into account these 3NFs 
except for the very backward angles.

All the deuteron analyzing powers, with exception of $T_{21}$, reveal  
at the highest energy 250 MeV/N and around  c.m.\  angles 
$\theta_{\rm c.m.}\gtrsim 120^\circ$ 
large  discrepancies to theory based on NN forces alone, which are
not resolved completely by the inclusion of the 3NFs.  
Such behavior of the deuteron analyzing powers is 
quite similar to that  
of the cross section and proton/neutron  analyzing powers
at 250 MeV/N found in Refs.~\cite{hatanaka02,maeda06}. 

%\\
The energy dependence of the predicted 3NF effects and 
the difference between the theory and the data 
for the deuteron analyzing powers is not always similar 
to that of the cross section and nucleon  analyzing power.
The vector analyzing power $iT_{11}$ 
and the tensor analyzing power $T_{20}$ 
have features similar to those of the cross section 
and the proton analyzing power $A_y^p$.
However the tensor analyzing power $T_{21}$ and $T_{22}$ reveal 
different energy dependence 
from that of  other observables.
Starting from $\sim 100$ MeV/N 
large 3NF effects are predicted. For
$T_{21}$ they are of different  magnitude for TM99 and Urbana IX and the
$T_{21}$ data seem to prefer the smaller effects of Urbana IX. 
For $T_{22}$ the large effects of TM99 and Urbana IX are practically the same.
At 200 MeV/N and below adding 3NFs worsens the description of data 
in a large angular region. 
It is contrary to what happens at the highest energy 
250 MeV/N,
where large 3NF effects are supported by the $T_{22}$ data 
in a large angular range.

The results obtained 
for the highest energy of 250 MeV/N indicate 
that some significant components 
are missing in the calculations, especially 
in the regions of higher momentum transfer. 
One possible candidate is relativistic effects.
We estimated their magnitude for the deuteron tensor analyzing powers 
by comparing nonrelativistic and relativistic predictions based on 
the CD Bonn potential~\cite{wit2005,wit2008}. 
They turned out to be small and only slightly 
alter the deuteron analyzing powers. 

%The smallness of the considered relativistic effects makes it appear
%that important parts of the 3NFs are missing. 
Due to the smallness of the considered relativistic effects,
it appears that important parts of the 3NFs are missing.
In the meson exchange picture used here, 
contributions of heavy mesons, e.g. $\pi$--$\rho$ and 
$\rho$--$\rho$ exchanges are omitted. 
The importance of such contributions is also seen in $\chi$EFT, 
where the leading non-vanishing 3NF consists of a $2\pi$-exchange, 
a $1\pi$ exchange-contact and a pure contact interaction topology. 
 With increasing energy orders of 
 the chiral expansion higher than N$^2$LO 
 become important, which introduce a multitude of additional
 short-range 3NF contributions, coming with different momentum-spin
 dependence. The complicated energy dependence of the deuteron
 analyzing powers and their strong dependence on the 3NF model used
 indicates the importance of such short-range components in their full
 description. The results of Faddeev calculations based on a force
 model with explicit $\Delta$ degrees of freedom show large changes in
 the predicted analyzing powers~\cite{deltuva} and also indicate 
 the importance of short-range 3NF components in the description of
 these spin observables.

In summary 
we have reported the first complete set of high precision data 
for the deuteron analyzing 
powers $iT_{11}$, $T_{20}$, $T_{21}$, and $T_{22}$,
in  elastic $dp$ scattering at 250 MeV/N, 
taken  in a wide angular range 
$\theta_{\rm c.m.} = 36^\circ$--$162^\circ$.
These data constitute a solid basis to guide  
theoretical investigations of 3NF models
at intermediate energies. 
It seems that missing short-range 3NF components are 
necessary in order to achieve a proper description 
of the data, such as e.g.\ provided by $\chi$EFT. 
So far the framework of the $\chi$EFT is only applicable up to 
$\sim$ 100 MeV/N for 3N scattering. 
It would be interesting, however, to see in future how converged 
theoretical predictions based on the $\chi$EFT forces describe our data.

%

%\begin{acknowledgments}
We acknowledge the outstanding work of the accelerator group
of RIKEN Nishina Center for delivering excellent polarized 
deuteron beams. 
We thank A. Yoshida, K. Kusaka, and T. Ohnishi for their strong 
support of the NP0702-RIBF23 
experiment.
This work was supported financially in part by the Grants-in-Aid 
for Scientific Research Numbers 20684010 of the Ministry of
Education, Culture, Sports, Science, and Technology of Japan.
It was also partially supported by the Polish 2008-2011 science 
funds as the research project No. N N202 077435 
and by the Helmholtz
Association through funds provided to the virtual institute ``Spin
and strong QCD''(VH-VI-231). 
The numerical calculations were performed on the 
supercomputer cluster of the JSC, J\"ulich, Germany.
%\end{acknowledgments}

%\newpage %Just because of unusual number of tables stacked at end
\bibliography{apssamp}% Produces the bibliography via BibTeX.

\begin{thebibliography}{}\label{sec:TeXbooks}

\bibitem{fuj57}
J. Fujita and H. Miyazawa, Prog.\ Theor.\ Phys. {\bf 17}, 360 (1957).
\bibitem{TM}
	S.~A.\ Coon and W. Gl\"ockle,  Phys.\ Rev.\ C{\bf 23}, 1790 (1981).
\bibitem{uIX}
	B.~S.\ Pudliner {\it et al.}, Phys.\ Rev.\ C{\bf 56}, 1720
	(1997).

\bibitem{kolk1994} U. van Kolck, Phys. Rev. C~{\bf 49}, 2932 (1994).

\bibitem{epel2006} 
E. Epelbaum, H.-W. Hammer, U.-G. Mei{\ss}ner, 
Rev. Mod. Phys. {\bf 81}, 1773 (2009).

\bibitem{AV18}
R.~B.~Wiringa {\it et al.},
	Phys.\ Rev.\ C {\bf 51}, 38 (1995).
%2
\bibitem{cdb}
	R.~Machleidt, Phys.\ Rev.\ C\ {\bf 63}, 024001 (2001).
%5
\bibitem{nijm}
V.~G.~J.~Stoks {\it et al.},
	Phys.\ Rev.\ C {\bf 49}, 2950 (1994).
%
\bibitem{chen} 
C. R. Chen {\it et al.}, Phys. Rev. C{\bf 33}, 1740 (1986).
\bibitem{sasakawa}
T. Sasakawa and S. Ishikawa, Few-Body Syst. {\bf 1}, 3 (1986).
%
\bibitem{underbind}
	A. Nogga {\it et al.},  Phys. Rev. C{\bf 65}, 054003 (2002).
%
%

%8
\bibitem{piep2002} 
%R.\ B.\ Wiringa {\it et al.},
%Phys.\ Rev.\ C {\bf 62}, 014001 (2000); 
S.\ C.\ Pieper {\it et al.}, Phys.\ Rev.\ C {\bf 66}, 044310 (2002).

%\bibitem{piep2001} 
%	S.\ C.\ Pieper {\it et al.},
%	Phys.\ Rev.\ C.\ {\bf 64}, 014001 (2001).
%
\bibitem{navr03} P. Navr\'atil and W.E. Ormand, Phys.\ Rev.\ C {\bf 68},
	034305 (2003).
%
%\bibitem{akmal98} see e.g. 
%A.\ Akmal {\it et al,}, Phys. Rec. C {\bf 58}, 1804 (1998).
%
\bibitem{wit98}
H. Wita{\l}a {\it et al.}, 
	Phys. Rev. Lett. {\bf 81}, 1183 (1998).
%
\bibitem{sak96} N.\ Sakamoto {\it et al.}, 
        Phys.\ Lett.\ B {\bf 367}, 60 (1996).
%42
\bibitem{sak2000}
H.\ Sakai {\it et al.}, Phys.\ Rev.\ Lett.\ {\bf 84}, 5288 (2000). 
%
\bibitem{sek2002}
K.\ Sekiguchi {\it et al.}, 
Phys.\ Rev.\ C\ {\bf 65}, 034003 (2002). 
%
\bibitem{hatanaka02}
K.\ Hatanaka {\it et al.}, 
Phys.\ Rev.\ C {\bf 66}, 044002 (2002).
%
%\bibitem{erm2003} 
%K.\ Ermisch {\it et al.}, Phys.\ Rev.\ C {\bf 68}, 051001 (2003).
\bibitem{erm2005} 
K.\ Ermisch {\it et al.}, Phys.\ Rev.\ C {\bf 71}, 064004 (2005).
%
\bibitem{sek2005}
K.\ Sekiguchi {\it et al.},
Phys.\ Rev.\ Lett. {\bf 95}, 0162301 (2005).
%
\bibitem{maeda06}
Y.\ Maeda {\it et al.}, 
Phys.\ Rev.\ C {\bf 76}, 014004 (2007).
%
\bibitem{anpow1}  
E.\ J.\ Stephenson {\it et al.}, Phys.\ Rev.\ C {\bf 60}, 061001 (1999).
\bibitem{anpow2}  
R.\ Bieber {\it et al.}, Phys.\ Rev.\ Lett.\ {\bf 84}, 606 (2000).
%\bibitem{anpow3}  
%K.\ Ermisch {\it et al.}, Phys.\ Rev.\ Lett.\ {\bf 86}, 5862 (2001). 
%
\bibitem{Przewoski_PRC_74_2006_064003}
B.\ v.\ Przewoski {\it et al.}, Phys.\ Rev.\ C {\bf 74}, 064003 (2006).
%
\bibitem{sek2004}
K.\ Sekiguchi {\it et al.},
Phys.\ Rev.\ C {\bf 70}, 014001 (2004).
%
\bibitem{hamid2007}
H. R. Amir-Ahmadi {\it et al.},
Phys.\ Rev.\ C {\bf 75}, 041001 (2007).
%
\bibitem{tm99} 
S.A. Coon, H.K. Han, Few Body Syst., {\bf 30}, 131 (2001).
%
\bibitem{deltuva2005}
A. Deltuva, A.C. Fonseca, P.U.Sauer, Phys. Rev. C\ {\bf 72}, 
054004 (2005).
%
%
\bibitem{Hossein_EPJA41_2007_383} 
H.\ Mardanpour {\it et al.}, Eur.\ Phys.\ J.\ A {\bf 31}, 383 (2007).
%
\bibitem{ohl72} 
G.\ G.\ Ohlsen, Rep.\ Prog.\ Phys.\ {\bf 35}, 717 (1972).
%
\bibitem{okamura95} 
H. Okamura {\it et al.}, AIP Conf. Proc. \textbf{343}, 123 (1995).
%
\bibitem{glo96}
W. Gl\"ockle {\it et al.}, Phys. Rep. {\bf 274}, 107 (1996).
%
\bibitem{hub97}
D. H\"uber {\it et al.}, Acta Phys. Polonica B {\bf 28}, 1677 (1997).
%

\bibitem{wit2005}
H. Wita{\l}a   {\it et al.}, 
                Phys.\ Rev.\ C {\bf 71}, 054001 (2005).

\bibitem{wit2008}
H. Wita{\l}a   {\it et al.}, 
                Phys.\ Rev.\ C {\bf 77}, 034004 (2008).


\bibitem{deltuva} A. Deltuva, private communication.

\end{thebibliography}

\end{document}